\documentclass[preprint]{aastex}

\usepackage{graphicx}

\newcommand{\be}{\begin{equation}}
\newcommand{\ee}{\end{equation}}
\newcommand{\ba}{\begin{eqnarray}}
\newcommand{\ea}{\end{eqnarray}}
\newcommand{\gr}{$\gamma$-ray }
\newcommand{\grs}{$\gamma$-rays }
\newcommand{\grsn}{$\gamma$-rays}
\newcommand{\gcs}{3EG~J1746-2851 }
\newcommand{\gcsn}{3EG~J1746-2851}
\newcommand{\TS}{{\rm TS }}
\newcommand{\gm}{G_{\rm Mult}}
\newcommand{\gb}{G_{\rm Bias}}
\begin{document}

\date{}  
\title{ A variability and localization study of 3EG J1746-2851}
\author{M. Pohl}
\affil{Department of Physics and Astronomy,
Iowa State University\\
Ames, Iowa 50011-3160, USA}
\altaffiltext{2}{mkp@iastate.edu}

\begin{abstract}
I have studied the variability properties 
and localization of \gcs based on EGRET data of the observing periods 1--4.
Using corrections for know systematic problems and performing various consistency checks I find 
no evidence of variability with an amplitude exceeding 30\%
with the possible exception of viewing period 429, for which a strong soft excess is observed. 
\gcs is displaced from the exact Galactic Center towards positive Galactic longitudes.
Sgr A$^\ast$, the center of Sgr A East, the pulsar J1747-2958,
and the TeV \gr source observed with HESS seem to be excluded as possible counterparts at the $> 95\%$ level.
\end{abstract}

\section{Introduction}

The Galactic Center region contains a many potential sources of high-energy \gr emission \citep{mf01}.
A point source of GeV-scale \gr emission, \gcsn, has been detected with EGRET \citep{hmh98}, but no firm identification
with sources known in other frequency bands has been made to date. Many models have been published to explain 
\gcsn, e.g. invoking an advection-dominated accretion flow onto the supermassive black hole known as Sgr~A$^\ast$
\citep{maha97,mms97}, the decay of neutralinos \citep{bere92}, the acceleration of quasi-monoenergetic electrons 
in the radio arc \citep{po97}, or a pulsar located on the line-of-sight to the Galactic Center \citep{mc03}.

Very high-energy \grs have also been detected from the Galactic Center region with the Whipple 10m telescope
\citep{kos04}, the CANGAROO telescope \citep{tsu04}, and the HESS array \citep{aha04}. The HESS data indicate
a power-law spectrum with photon index $\alpha=2.21\pm 0.09$, an extrapolation of which does not fit the GeV-scale 
flux measured with EGRET. The source of TeV-scale \gr emission shows no evidence of variability and is coincident
with Sgr A$^\ast$ to within about 0.02$^\circ$. The lack of a cut-off in the \gr spectrum seems to disfavor models
involving the annihilation of dark-matter particles \citep{ber98,ell02,gp03}, for a very high mass $m_x \gtrsim 12$~TeV
of the hypothetical neutralino would be required.

More conventional models for the high-energy emission from the Galactic Center region have also been proposed.
Electron acceleration by plasma turbulence near the supermassive black hole \citep{sch02} may be a very efficient
process \citep{lpm04}.
\citet{ad04} argue that the combination of ADAF-type accretion with a subrelativistic MHD outflow can explain
the TeV-scale emission as well as that at X-rays and lower frequencies, but \gcs would be a separate source.
\citet{fm03} discuss Sgr A East, a nonthermal radio source with a supernova-like morphology located near the 
Galactic center, as a possible counterpart of \gcsn, and \citet{cro04} attempt to relate \gcs, 
the TeV-scale \gr source, and anisotropies in the UHE cosmic-ray intensity to Sgr A East.

One of the fundamental questions is whether or not the sources seen in the GeV and TeV bands are identical.
The analysis of EGRET data for the Galactic Center region is known to be affected by systematic problems, some
of which have been understood since the detection of \gcs was announced in 1998. In this paper I
re-analyze the available EGRET data with a view to infer the variability properties and the localization as
accurately as reasonably possible. For that purpose I introduce corrections in the way the diffuse foreground
model is treated in the likelihood analysis, which are
outlined in Sec.~\ref{foreground}. I do not change the foreground model per se. Various consistency
checks are performed using data for different energy bands and from different observing dates. 

\section{Modeling the galactic foreground emission}
\label{foreground}
I have first analyzed the combined data sets of the EGRET observing periods 1--4 in an area 
of $\pm 60^\circ$ in latitude and longitude around the Galactic Center. Using all 80 sources, that are 
listed in the Third EGRET Catalogue \citep{hart99} to be located in an area  
of $\pm 40^\circ$ in latitude and longitude around the Galactic Center, I have run the LIKE 
software \citep{mattox96} to simultaneously determine the multipliers $\gm$ and $\gb$ for the
standard model of galactic diffuse
\gr emission \citep{hunter97} and isotropic emission, respectively, and
the fluxes of these 80 sources while keeping their 
positions fixed.
The multipliers $\gm$ and $\gb$ were allowed to vary freely. The purpose of the exercise was two-fold:
\begin{itemize}
\item to test whether or not the combination of the eighty 3EG sources and the foreground model 
is an acceptable fit to the entire data set. If there are \gr excesses with test statistic 
$\TS \ge 16$ within 30$^\circ$ of the Galactic Center, we may need to incorporate them in the 
analysis of individual viewing periods.
\item to derive an appropriate range of parameter values for the foreground model. The galactic diffuse
emission cannot be variable, and therefore the same model, i.e. the same values for $\gm$ and $\gb$
that we obtain from analyzing the combined data set,
should be used in the likelihood analysis of individual viewing periods. However,
$\gm$ and $\gb$ must be allowed
to vary within the range of statistical uncertainty of their best-fit values and within the uncertainty
range of the spark-chamber efficiency correction \citep{esp99}.
\end{itemize}
In contrast to the procedure used to derive the Third EGRET Catalogue,
I did not include below-threshold excesses with $\TS \ge 9$ in the analysis. The source fluxes obtained
for the combined data set of the EGRET observing periods 1--4 are generally very similar to those
listed in \citet{hart99}. The most prominent exception is 3EG~J1744-3011, located 
at $l=-1.15^\circ$ and $b=-0.52^\circ$, with
\ba
3~{\rm EG}&\qquad S=9.4\,\sigma&\qquad F(> 100\,{\rm MeV})=(63.9\pm 7.1)\,10^{-8}\,{\rm ph./cm^2/sec}
\nonumber\\
{\rm This\ analysis}&\qquad S=7.6\,\sigma&\qquad F(> 100\,{\rm MeV})=
(50.5\pm 6.9)\,10^{-8}\,{\rm ph./cm^2/sec}
\ea
For the Galactic Center source, 3EG~J1746-2851, the flux and source significance determined with the
combined data set are virtually identical to those listed in the catalogue.
\ba
3~{\rm EG}&\qquad S=17.5\,\sigma&\qquad F(> 100\,{\rm MeV})=(119.9\pm 7.4)\,10^{-8}\,{\rm ph./cm^2/sec}
\nonumber\\
{\rm This\ analysis}&\qquad S=17.5\,\sigma&\qquad F(> 100\,{\rm MeV})=
(118.6\pm 7.3)\,10^{-8}\,{\rm ph./cm^2/sec}
\label{2}
\ea
At the position of the Galactic Center source the best-fit parameters of the foreground
model are
\be 
{\gm}=0.948\qquad\qquad {\gb}=3.405
\label{3}
\ee
In the entire area of $\pm 20^\circ$ in latitude and longitude around the Galactic center
these parameters have values not exceeding the range $[0.93,1.18]$ for $\gm$ and
$[2.,4.]$ for $\gb$. The high value for the multiplier of isotropic emission, $\gb$, is possibly
related to inverse-Compton emission not accounted for in the standard model of galactic diffuse
\gr emission.

The highest intensity of diffuse galactic \grs is observed in direction of the Galactic Center.
Any systematic errors associated with a mismodeling of the galactic emission will therefore be 
most severe in this region of sky. Systematic effects will also be more important at low \gr energies 
than at high energies on account of the strong energy dependence of EGRET's point-spread
function (PSF). Given the average photon flux above 100 MeV (Eq.\ref{2}) and
the similarity between the apparent \gr spectrum of \gcs \citep{mer96} and
that of the diffuse emission, we can infer that the mean measured
intensity (or count density) associated with \gcs within the solid angle
element containing 68\% of the photons, $\Omega (68\,\%)$, is 
approximately
\be
I_{\rm GC} (68\,\%) \simeq 0.68 \,{{F(> 100\,{\rm MeV})}\over {\Omega (68\,\%)}}
\simeq 1.8\cdot 10^{-4}\ {\rm ph./cm^2/sec/sr}
\label{4}
\ee
which is small compared with the mean intensity of diffuse emission in that solid angle element.
In the energy band above 1 GeV the mean 
intensity associated with \gcs is actually more than a factor of two higher than that of the 
diffuse emission, solely
on account of the narrower PSF. We must therefore espect that the high-energy band 
above 1 GeV is relatively clean 
of systematic problems, as far as the determination of flux and position of \gcs are concerned, 
and that the low-energy bands and the very wide bands, e.g. $>$ 100 MeV, are subject to possibly 
serious systematic uncertainties.

One such systematic uncertainty arises from a simplification in the EGRET standard data 
analysis chain. There one uses allsky maps of modeled galactic diffuse \gr emission, in which the effect
of the PSF is already incorporated. These maps are then multiplied with the exposure map (the
time integral of the effective area) of the
viewing period in question to produce a map of the expected event distribution that can be compared
with the data. So the PSF is applied before the exposure, whereas it should be the other way around.
Unpublished earlier Monte-Carlo studies by the author have indicated that this simplification may be
problematic for \gcsn, if the pointing direction is in the Galactic Plane at 
galactic longitude $l\approx \pm 20^\circ$, so that the Galactic Plane is radially oriented
in the EGRET field-of-view, and the effective area of the instrument is
rapidly decreasing with off-axis angle at the location of the point source in question.

For each viewing period I have created customized maps of diffuse galactic \gr emission, that
are corrected for this simplification. Using a simple analytical representation of the spectrum of the
diffuse galactic \gr emission,
\be
I_{\rm dif}\propto \left\{ \begin{array}
{r@{\qquad{\rm for}\quad}l}
E^{-1.6} & E < 1\ {\rm GeV}\\
E^{-2.3} & E > 1\ {\rm GeV}
\end{array} \right.
\ee
as well as the effective area distribution, the
energy dispersion function, and the tabulated PSF data from the pre-launch calibration files for the
dominant EGRET observing mode 74, I have calculated a mean PSF for each energy interval, for which 
data analysis would be performed. The mean PSF thus derived was then applied to the expected true count 
distribution, i.e. the product of the model of galactic emission, GALDIF, and the exposure map,
EX, to derive the expected count distribution in the data,
\be 
C(\theta,\phi)= \int d\alpha\,d\beta \ \ {\rm PSF}(\theta,\phi,\alpha,\beta)\ {\rm EX}(\alpha,\beta)\ 
{\rm GALDIF}(\alpha,\beta)
\ee
where $\alpha$ and $\beta$ denote the true angular coordinates of the incoming \grsn,
whereas $\theta$ and $\phi$ are the measured angular coordinates of the \grsn. The standard maps 
of diffuse galactic \gr emission, that can be obtained from the CGRO
Science Support Center, are calculated assuming a single power-law spectrum with index
2.1, thus implying a PSF of different width that used in my analysis.

\section{Analysis}
Table \ref{t1} displays a list of all viewing periods considered in this study. For each viewing period
the data have been analyzed in the energy bands 100-300~MeV, 300-1.000~MeV, $>$100~MeV, $>$300~MeV,
and $>$1.000~MeV. I have also constructed a combined data set for all viewing periods with a view to 
derive a detailed spectrum and to pinpoint the localization of the source.

\begin{deluxetable}{rrrr}
\tablecolumns{4}
\tablewidth{0pc} 
\tablecaption{A list of all viewing periods considered in this study. The table also 
shows the pointing direction in galactic coordinates and the observing time.}
\tablehead{\colhead{VP}&\colhead{$l$}&\colhead{$b$} &\colhead{t [$10^5$\ sec]}}
\startdata
5&0.0&-4.0&11.9\\
16&0.0&20.3&12.9\\
210&-4.4&6.3&2.4\\
214&-4.4&6.3&2.5\\
219&-9.9&15.9&1.1\\
223&-0.9&-0.1&2.6\\
226&-5.0&5.0&8.6\\
2295&5.0&5.0&4.2\\
231&22.2&-13.1&5.7\\
232&-12.5&0.0&12.1\\
3023&1.4&9.3&10.3\\
323&-3.2&-11.3&11.9\\
324&15.0&5.6&5.9\\
330&18.0&0.0&3.4\\
332&18.0&0.0&14.6\\
334&9.0&-8.4&5.9\\
3365&-19.6&2.9&4.5\\
421&-4.7&0.4&5.8\\
422&-4.6&-0.4&6.1\\
423&2.6&-0.2&8.5\\
4235&-14.3&13.5&8.5\\
429&18.3&4.0&6.0\\
508&6.5&-0.2&5.1\\
625&1.5&0.8&11.1\\
\enddata
\label{t1}
\end{deluxetable}

\subsection{Consistency checks and the spectrum}

\begin{figure}
\plotone{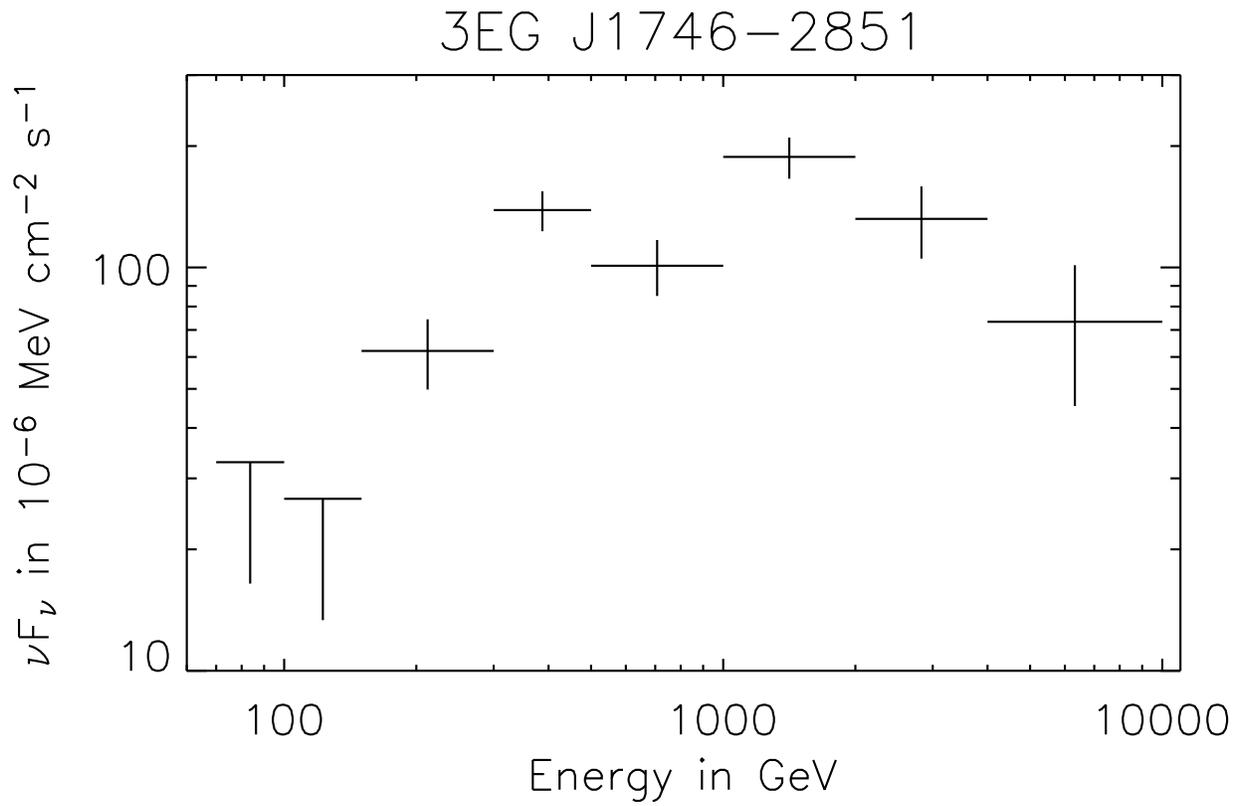}
\caption{The $\nu\,F_\nu$ spectrum of \gcs as derived from an analysis of the combined data as
listed in table 1. The spectrum is similar to that of the galactic diffuse emission.}
\label{f1}
\end{figure}

The spectra of EGRET sources are usually determined on the basis of a likelihood analysis in ten
standard energy ranges. Table~\ref{t2} lists the integrated photon flux and the
$\nu\,F_\nu$ flux in the ten narrow energy bands and five wide energy bands based on a likelihood
analysis of the combined data set. The spectrum depends very weakly on the assumed position of the source,
if that is varied by $\delta l \lesssim 0.1^\circ$ in the Galactic Plane. 
Figure~\ref{f1} shows the $\nu\,F_\nu$ spectrum above 70 MeV.
The parameters of the diffuse foreground model,
$\gm$ and $\gb$, were allowed to vary freely. Table~\ref{t2} also lists the integrated photon flux
for all energy bands that one obtains using the standard maps of diffuse foreground emission.
The combined data set has an exposure distribution that peaks within 2$^\circ$ of the Galactic Center
and is fairly symmetric. The standard maps and the modified maps of galactic diffuse emission
differ only marginally.

The photon flux for the energy bands below 100~MeV is strongly
dependent on the particulars of the likelihood run, e.g. on the analysis parameter RANAL, and thus
discredited. An anticorrelation between the multiplier $\gm$ of the diffuse emission model
and the photon flux attributed to \gcs can be observed, which indicates a similarity of the expected
angular distributions of events for the diffuse emission and \gcsn , respectively. 
The results of the likelihood analysis at energies below 100~MeV do therefore not merit our consideration
in the variability and localization study.

One also notes that the integrated flux above 100 MeV is significantly higher than derived in the 
allsky analysis (Eq.\ref{2}). The only difference between the two likelihood analyses was the inclusion
in the allsky analysis of all viewing periods not listed in table 1, which leads to a significantly more
homogeneous distribution of exposure across the sky. However, the exposure distribution for the combined
data set as listed in table 1 is wider than that of any individual viewing period.
So if the likelihood analysis of the combined data set is apparently inaccurate at low
\gr energies for which the PSF is wide, there is no reason to assume it would fare better for any
individual viewing period.

\begin{deluxetable}{rrrr}
\tablecolumns{4}
\tablewidth{0pc} 
\tablecaption{The spectrum of \gcs as derived using the combined data set. The integrated photon flux 
in each energy band is given in units of $10^{-8}\ {\rm cm^{-2}\, s^{-1}}$. The $\nu\,F_\nu$ flux 
is in units of $10^{-6}\ {\rm MeV\, cm^{-2}\, s^{-1}}$ and derived assuming a photon
spectrum $\propto E^{-2}$. The results of the likelihood analysis at 
energies below 70~MeV should be considered as upper limits for systematic reasons. Upper limits are
given for a 2$\sigma$ confidence level. For comparison the fourth column gives the 
integrated photon flux derived using the standard maps of diffuse emission.}
\tablehead{\colhead{Energy [MeV]}&\colhead{$\overline{F}$}&\colhead{$\nu\,F_\nu$}
&\colhead{$\overline{F}_{\rm strd}$}}
\startdata
30$-$50&565.3$\pm$73.1&424.0$\pm$54.8&344.9$\pm$71.6\\ 
50$-$70&101.8$\pm$16.9&178.2$\pm$29.6&73.4$\pm$16.7\\
70$-$100&$<$14.1&$<$32.6&$<$15.4\\
100$-$150&$<$8.92&$<$26.3&$<$9.1\\
150$-$300&20.7$\pm$4.1&59.7$\pm$12.3&16.1$\pm$4.1\\
300$-$500&18.5$\pm$2.1&136.$\pm$15.8&17.9$\pm$2.1\\
500$-$1.000&10.1$\pm$1.6&97.$\pm$16.&9.7$\pm$1.6\\
1.000$-$2000&9.4$\pm$1.1&181.$\pm$22.&9.1$\pm$1.1\\
2000$-$4000&3.3$\pm$0.67&127.$\pm$26.8&3.2$\pm$0.67\\
4000$-$10.000&1.1$\pm$0.42&68.$\pm$28.&1.1$\pm$0.42\\
100$-$300&22.4$\pm$6.5&33.6$\pm$9.8&12.6$\pm$6.5\\
100$-$10.000&143.7$\pm$7.0&145.$\pm$7.1&122.1$\pm$7.3\\
300$-$1.000&27.5$\pm$2.6&118.$\pm$11.1&25.8$\pm$2.6\\
300$-$10.000&48.5$\pm$3.1&150.$\pm$9.6&45.7$\pm$3.1\\
1.000$-$10.000&14.5$\pm$1.5&161.$\pm$16.7&13.7$\pm$1.5\\
\enddata
\label{t2}
\end{deluxetable}

Another consistency check would be a comparison of the integrated flux obtained for a wide energy band
with the sum of the fluxes in the corresponding narrow energy bands. While the high-energy bands are
consistent with each other, we note that for the wide band 100--10.000 MeV the flux is about twice
the sum of fluxes obtained in the corresponding narrow energy bands. The flux determined for
the 300--10.000 MeV band is about 14\% higher than the sum of the fluxes for the corresponding narrow
energy bands. The chance probability of that deviation is about 15\%, so given the number of trials
we have to expect one deviation of that amplitude on statistical grounds. Significant problems
can thus only be noted for the energy band 100--10.000 MeV.

\subsection{Search for variability}
\label{s1}
In the preceding subsection we have reported on a number of inconsistencies in the results of a
likelihood analysis of the combined data set, which indicate systematic problems. The affected 
data in the energy bands 100--10.000 MeV and below 100 MeV will therefore be excluded from the 
following analysis. 

We now want to analyze the data for individual viewing periods with a view to infer the
variability properties of \gcsn. Earlier publications \citep[e.g.][]{nol03}
reported some evidence for variability based on 
a likelihood analysis of EGRET data in the energy band 100--10.000 MeV that we found affected by 
systematic problems. Here we use only data for energy bands, for which we see no clear evidence for 
systematic errors.

In the search for variability we therefore analyze EGRET data in the energy bands
100--300 MeV, 300--1.000 MeV, 1.000--10.000 MeV, and 300--10.000 MeV. In the narrow energy bands
that we have used to derive the spectrum of \gcs the statistical significance of detection
is so low that an analysis of data of individual viewing periods does not provide meaningful results.
We keep the position of the source fixed, for that cannot vary. Our analysis of the combined
data set yielded $l=0.11^\circ$ and $b=-0.04^\circ$ as best-fit position of \gcsn, which we henceforth
use as fixed location in the search for variability. Small changes in the assumed position
by $\delta l \lesssim 0.1^\circ$ have virtually no impact on the results of the variability analysis.

The intensity of diffuse emission should be constant over time, and so should be the values of 
its scaling factors in the likelihood analysis, $\gm$ and $\gb$. The uncertainty of the spark-chamber
efficiency correction introduces some variations in the best-fit values of the scaling factors, that
should have an amplitude around 10\% \citep{esp99}. For each data set we ran the likelihood analysis 
twice, first with the scaling factors fixed to the values found in the allsky analysis (Eq.~\ref{3}),
and then allowing the scaling factor to freely vary. Variable scaling factors should provide a better
representation of the diffuse galactic emission, for they can balance errors in the spark-chamber
efficiency correction and they can correct inappropriate large-scale structure in the model of
diffuse emission. On the other hand, one may expect an anti-correlation between the best-fit
flux from \gcs and the best-fit value of $\gm$, which is the dominant parameter for the intensity
of the expected diffuse \gr emission from the inner galaxy. Therefore
the mean relative amplitude of variations in $\gm$ over $N$ viewing periods
\be
\sigma{\gm}=\sqrt{{1\over N}\,\sum_{i=1}^N\ {{(\gm-\overline{\gm})^2}\over {\overline{\gm}^2}}}
\label{7}
\ee
should not significantly exceed 10\%. We thus monitor the variations in the best-fit values of
$\gm$ to ensure that variations in the measured flux from \gcs are not caused by an unrealistically
extreme value of $\gm$.

\subsubsection{The energy band 300--10.000 MeV}

The results of the likelihood analysis runs for the individual viewing periods as well as for the
combined data set are summarized in table \ref{t3}. The scaling parameter of diffuse emission,
$\gm$, has best-fit values that are characterized by
\be 
\overline{\gm}=1.021\qquad  \sigma{\gm}=0.106
\label{8}
\ee
so the scatter in $\gm$ is at a level commensurate with the systematic error in the absolute
determination of EGRET's effective area at the time of measurement, thus not indicating any additional
systematic problem.

\begin{deluxetable}{rrrrrr}
\tablecolumns{6}
\tablewidth{0pc} 
\tablecaption{Results of the likelihood analysis in the energy
band $>\,$300 MeV. The position of \gcs was held fixed at $l=0.11\deg$ and $b=-0.04\deg$.
The integrated photon flux $F$ is given in units of 
$10^{-8}\  {\rm cm^{-2}\, s^{-1}}$. Upper limits correspond to 95\%
confidence or 2$\sigma$.}
\tablehead{\colhead{ }&\multicolumn{2}{|c|}{$\gm$ fixed}&\multicolumn{3}{|c|}{$\gm$ variable}\\
\colhead{VP}&\colhead{$\sigma$}&\colhead{$F$}&\colhead{$\sigma$}&\colhead{$F$}
&\colhead{$\gm$}}
\startdata
5&5.2&28.9$\pm$6.2&7.3&44.9$\pm$7.0&0.889\\
16&8.9&85.5$\pm$11.7&7.3&76.4$\pm$12.3&1.129\\
210&2.5&58.4$\pm$27.4&1.9&49.0$\pm$28.6&1.129\\
214&1.6&29.3$\pm$20.1&1.5&28.8$\pm$21.3&1.059\\
219&0.7&$<$125.7&0.8&$<$136.6&0.988\\
223&2.3&41.8$\pm$20.9&1.9&37.7$\pm$22.0&1.117\\
226&1.9&19.9$\pm$11.5&2.2&25.9$\pm$12.6&0.969\\
2295&2.1&32.8$\pm$17.6&2.1&37.1$\pm$19.4&1.015\\
231&2.4&56.6$\pm$29.0&2.4&62.6$\pm$31.2&0.980\\
232&5.6&61.8$\pm$13.0&5.4&64.5$\pm$14.0&1.029\\
3023&2.4&29.7$\pm$13.4&1.9&24.7$\pm$14.3&1.068\\
323&4.9&45.9$\pm$10.7&3.9&39.4$\pm$11.3&1.122\\
324&3.6&54.0$\pm$17.5&3.2&52.0$\pm$18.5&1.069\\
330&5.0&118.2$\pm$29.6&4.3&110.5$\pm$31.1&1.086\\
332&6.8&69.1$\pm$12.1&6.6&72.2$\pm$12.9&1.022\\
334&2.0&27.7$\pm$14.9&1.9&27.5$\pm$16.0&1.053\\
3365&2.4&55.2$\pm$26.9&1.9&46.3$\pm$27.9&1.120\\
421&1.8&20.5$\pm$12.7&2.5&32.2$\pm$14.5&0.938\\
422&1.2&12.9$\pm$11.2&3.2&38.1$\pm$13.5&0.822\\
423&1.9&17.3$\pm$9.8&3.7&38.1$\pm$11.6&0.897\\
4235&0.3& $<$ 37.2&2.2&34.2$\pm$18.0&0.773\\
429&3.2&56.7$\pm$20.7&4.0&77.8$\pm$23.3&0.883\\
508&5.6&88.7$\pm$19.7&4.0&69.5$\pm$20.2&1.235\\
625&4.2&43.1$\pm$11.7&3.2&36.4$\pm$12.6&1.107\\
all&17.9&44.7$\pm$2.9&17.8&48.5$\pm$3.1&1.011\\
\enddata
\label{t3}
\end{deluxetable}

We can perform a $\chi^2$-test to determine how well the measured fluxes in the 24 viewing
periods are compatible with a constant flux. The best-fit constant flux would be
\be
\overline{F} (> 300\, {\rm MeV}) = 45.74\cdot 10^{-8}\  {\rm cm^{-2}\, s^{-1}}
\label{9}
\ee
for which the minimal value of the $\chi^2$-sum and the chance probability of drawing the measured 
distribution given a constant flux, the goodness-of-fit, would be
\be
\chi_{\rm min}^2=30.0\qquad
P_{\rm chance}=15\%
\label{10}
\ee
This indicates some scatter in the distribution
of flux values beyond what must be expected on statistical grounds. We can estimate the amplitude of that
scatter by Gaussian adding of a systematic uncertainty $\delta = y\,\overline{F}$ to the flux
measurement error. The amplitude of scatter is then approximately given by the value of $y$, for which
the minimal value of the $\chi^2$-sum equals the number of degrees of freedom (22 in this test).
The resulting estimate for the amplitude of variations is 
\be
y_{\rm best\ fit} = 0.197
\label{11}
\ee
which is just twice the expected level of absolute calibration error.

\subsubsection{The energy band 300--1.000 MeV}

This energy band is part of the energy band, the results for which we have just discussed. It is thus
not statistically independent. The scaling parameter of diffuse emission,
$\gm$, has best-fit values that are characterized by
\be 
\overline{\gm}=0.875\qquad  \sigma{\gm}=0.118
\label{12}
\ee
so the scatter in $\gm$ is again at a level commensurate with the systematic error in the absolute
calibration.

The best-fit constant flux would be
\be
\overline{F} (300-1.000\, {\rm MeV}) = 25.08\cdot 10^{-8}\  {\rm cm^{-2}\, s^{-1}}
\label{13}
\ee
for which the minimal value of the $\chi^2$-sum and the chance probability of drawing the measured 
distribution given a constant flux, the goodness-of-fit, would be
\be
\chi_{\rm min}^2=31.0\qquad
P_{\rm chance}=12.3\%\ ,
\label{14}
\ee
very similar to the results obtained for the wide energy band 300--10.000 MeV.

The resulting estimate for the amplitude of variations is 
\be
y_{\rm best\ fit} = 0.346\ ,
\label{15}
\ee
which is nearly twice as much as in the wide energy band.

\subsubsection{The energy band 1.000--10.000 MeV}

At these very high energies we expect the model of diffuse galactic emission to play 
a small role in the
likelihood analysis. On the other hand, the small number of photons will reduce 
the statistical accuracy of the results.

The scaling parameter of diffuse emission,
$\gm$, has best-fit values that are characterized by
\be 
\overline{\gm}=1.46\qquad  \sigma{\gm}=0.154
\label{16}
\ee
so the scatter in $\gm$ is slightly larger than the systematic error in the absolute
calibration. The large value of $\overline{\gm}$ reflects the GeV excess in the 
diffuse galactic \gr emission \citep{hunter97}.

The best-fit constant flux would be
\be
\overline{F} (> 1.000\, {\rm MeV}) = 12.9\cdot 10^{-8}\  {\rm cm^{-2}\, s^{-1}}
\label{17}
\ee
for which the minimal value of the $\chi^2$-sum and the chance probability of drawing the measured 
distribution given a constant flux, the goodness-of-fit, would be
\be
\chi_{\rm min}^2=25.3\qquad
P_{\rm chance}=33.5\%\ ,
\label{18}
\ee
indicating no evidence for variability at energies higher than 1 GeV.

The resulting estimate for the amplitude of variations is 
\be
y_{\rm best\ fit} = 0.182\ ,
\label{19}
\ee
which here is more an estimate for the sensitivity of the method used in the search for
variability, for additional fluctuations witha an amplitude much higher than this value would be required 
to significantly increase the minimum $\chi^2$-sum in Eq.~\ref{18}.

\subsubsection{The energy range 100--300 MeV}

This data in this energy range should contain a large number of \gr events, but are potentially
problematic because of the large width of the PSF. Systematic problems were evident at lower 
energies, and we do not know very well to what extent the data at 100--300 MeV are affected.

The scaling parameter of diffuse emission,
$\gm$, has best-fit values that are characterized by
\be 
\overline{\gm}=0.897\qquad  \sigma{\gm}=0.132
\label{20}
\ee
so the scatter in $\gm$ is marginally larger than the systematic error in the absolute
calibration. 

The best-fit constant flux would be
\be
\overline{F} (100-300\, {\rm MeV}) = 29.9\cdot 10^{-8}\  {\rm cm^{-2}\, s^{-1}}
\label{21}
\ee
for which the minimal value of the $\chi^2$ sum and the chance probability of drawing the measured 
distribution given a constant flux, the goodness-of-fit, would be
\be
\chi_{\rm min}^2=43.58\qquad
P_{\rm chance}=0.6\%\ .
\label{22}
\ee
The resulting estimate for the amplitude of variations is 
\be
y_{\rm best\ fit} = 0.71\ .
\label{23}
\ee
A close inspection of the flux distribution reveals that all this evidence of variability is caused 
by one data point, that for viewing period 429. If we neglected that measurement, our 
results would be
\be
\overline{F} ({\rm excl.\ VP429}) = 26.1\cdot 10^{-8}\  {\rm cm^{-2}\, s^{-1}}
\label{24}
\ee
and
\be
\chi_{\rm min}^2=21.92\qquad
P_{\rm chance}=46.5\%\ \qquad {\rm excl.\ VP429}, 
\label{25}
\ee
i.e. no evidence for any variability. So the question is whether or not the measurement of a
high flux during VP429 is realistic.

\begin{figure}
\plotone{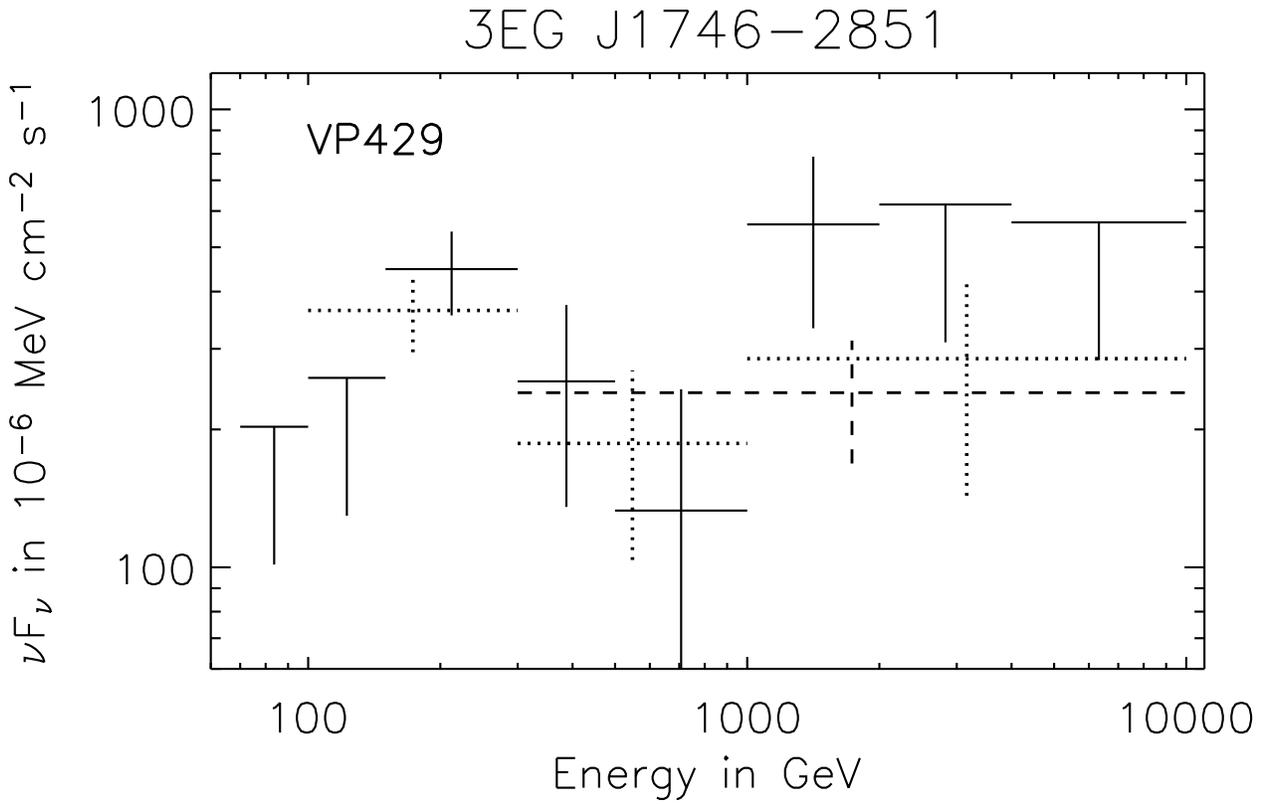}
\caption{The $\nu\,F_\nu$ spectrum of \gcs during viewing period 429. 
Data shown with dotted or dashed error bars refer to wide energy bands and 
are not statistically independent of the data points plotted with solid error bars.
The spectrum is much softer than the time-averaged spectrum shown in Fig.~\ref{f1}.}
\label{f2}
\end{figure}

The measured flux from \gcs during VP429 is
\be
F_{\rm VP429}(100-300\, {\rm MeV})=(242.6\pm 46.1) \cdot 10^{-8}\  {\rm cm^{-2}\, s^{-1}}
\label{26}
\ee
What was the flux in the other energy bands? In all cases it was around 60\% above average with
the statistical significance of the excess being around 1$\sigma$. Equation \ref{26} indicates that
in the energy band 100--300 MeV the flux was a factor of 8 above average with a 
statistical significance of 4.6$\sigma$. If the outburst were real, the source must have 
an soft flare spectrum, as shown in figure~\ref{f2}.

What was the best-fit value of the scaling factor of diffuse emission, $\gm$, for VP429?
It is $\gm ({\rm VP429})=0.66$, actually the lowest value of all viewing periods, 
which mandates a view at the results of the likelihood analysis for fixed $\gm$,
\be
F_{\rm VP429}(\gm=0.948) = (104.1\pm 39.9)\cdot 10^{-8}\  {\rm cm^{-2}\, s^{-1}}
\label{27}
\ee
which is still more than three times the average flux with a statistical significance
of 2$\sigma$.

\subsection{The localization of \gcs}

The likelihood ratio test, that is used to determine the flux and detection significance of point
sources, can also provide an estimate of the source position and its uncertainty. For that purpose
one calculates the point source test statistic
\be
\TS = -2\left[\ln L_0 - \ln L_1 (\alpha,\beta)\right]
\label{28}
\ee for a finely sampled array of test positions $(\alpha,\beta)$ \citep{mattox96}.
Here $L_0$ is the likelihood function for the null hypothesis and $L_1$ is the likelihood function
assuming a point source at the position $(\alpha,\beta)$. The increase in \TS at the 
location of its maximum 
over \TS at the true position is expected to follow a $\chi_2^2$ distribution, for the two angular
coordinates are two additional degrees of freedom \citep{wilks}. Decrements from $\max(\TS)$ 
by 2.3, 6.0, and 9.1 delineate the boundaries of regions in which the source is located with
a confidence level of 68\%, 95\%, and 99\%. 

Obviously the localization analysis can provide meaningful results only if $\max(\TS)$ is 
not too small. We have therefore first analyzed the combined data set, which should provide the
best statistical accuracy. As in our search for variability (Sec.~\ref{s1}), we exclude data
at energies below 100 MeV and in the wide energy band 100-10.000 MeV, because they appear 
to be affected by systematic problems. 

In the energy band 100--300 MeV \gcs is detected with
a significance of only 3.5$\sigma$, too low to provide constraining results, so here we only report
that the localization estimate for that energy band is compatible with those obtained using data
at higher photon energies.

\begin{figure}
\plotone{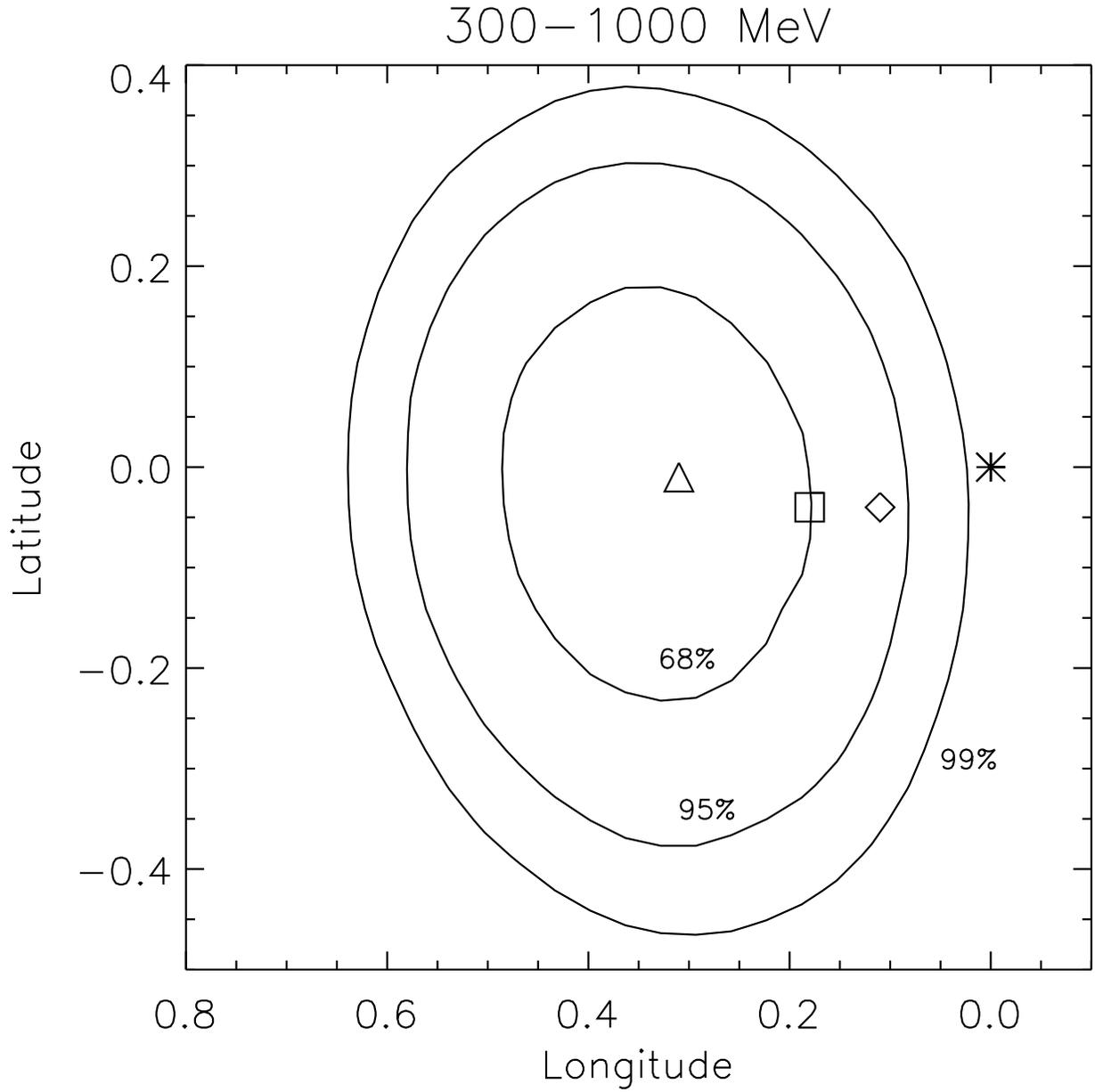}
\caption{Source localization confidence contours for the energy band 300--1.000 MeV. The
square indicates the position of the central part of the Galactic Center arc and the asterisk
denotes Sgr A$^\ast$. The triangle and the diamond indicate the best-fit position in the energy
bands 300--1.000 MeV and $>$ 1.000 MeV, respectively.}
\label{f3}
\end{figure}

\begin{figure}
\plotone{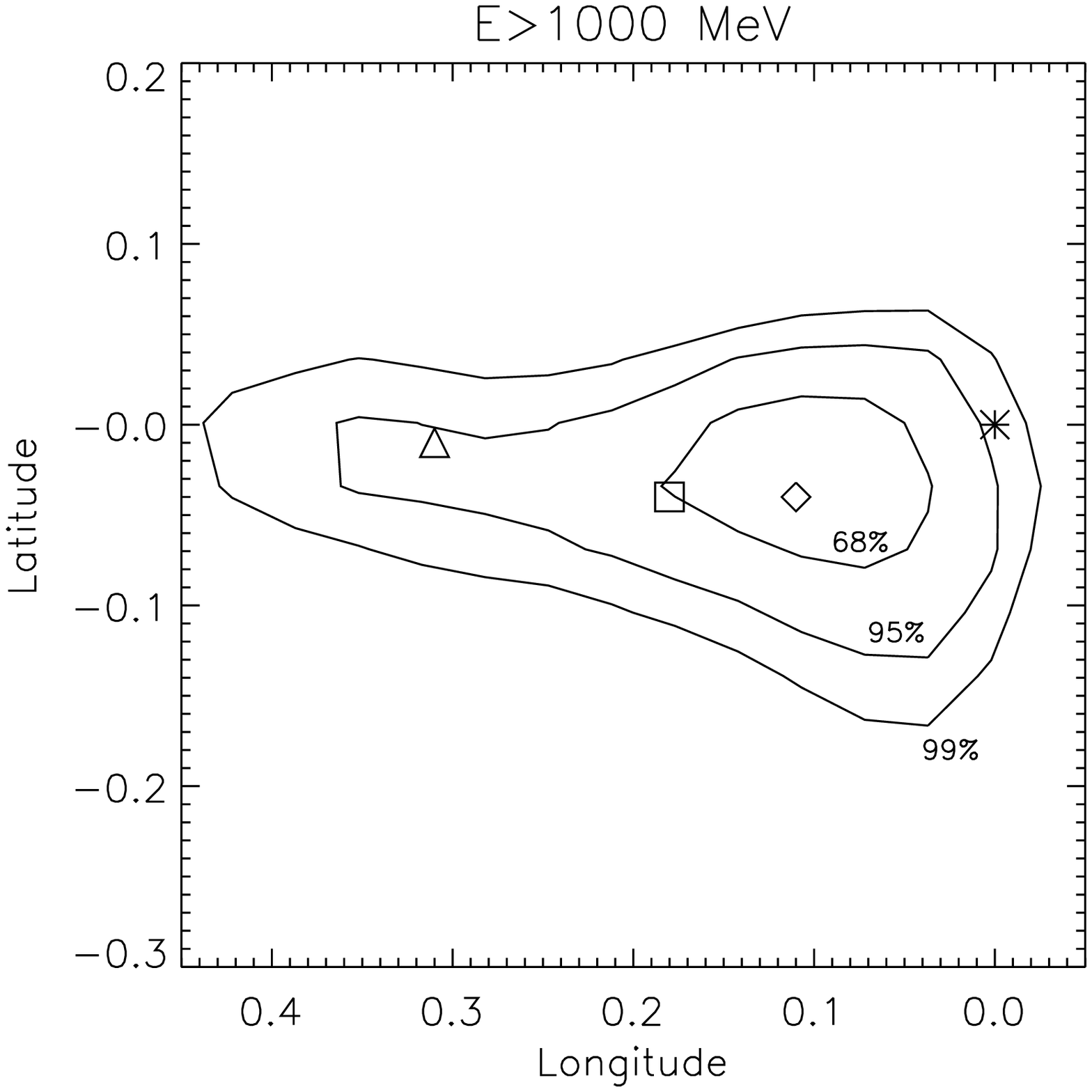}
\caption{The results of the source localization analysis for the energy band $>$ 1.000 MeV. The
symbols indicate the positions of the same systems as in Fig.~\ref{f3}.}
\label{f4}
\end{figure}

In the figures \ref{f3} and \ref{f4} we show source localization confidence contours for the energy 
bands 300--1.000 MeV and $>$ 1.000 MeV, respectively, in comparison with 
the position of the central part of the galactic center arc and that of
Sgr A$^\ast$. To be noted from the figures is that in the two independent
energy bands the likelihood analysis indicates a source position that is significantly 
displaced from the 
exact Galactic Center, or Sgr A$^\ast$, or the position of the TeV \gr source \citep{aha04},
that coincides with Sgr A$^\ast$ to within 1.2 minutes of arc.

The best-fit positions determined for the two energy bands are moderately well compatible with
each other. However, both are within the 68\% confidence contour as determined for the
100--300 MeV band. The 68\% confidence contours for the energy 
bands 300--1.000 MeV and $>$ 1.000 MeV just touch each other, but there is a substantial overlap 
of the 95\% confidence regions. The localization analysis of these two statistically independent
data sets thus provides a much more consistent result than did the original study by the EGRET
team \citep{hmh98}, where even the 99\% confidence contours for the 1--30 GeV and the 100--300
MeV bands didn't come close to each other.

We can further test for systematic errors in two ways, which we discuss in turn.
\subsubsection{Consistency of localization in different energy bands}
\begin{figure}
\plotone{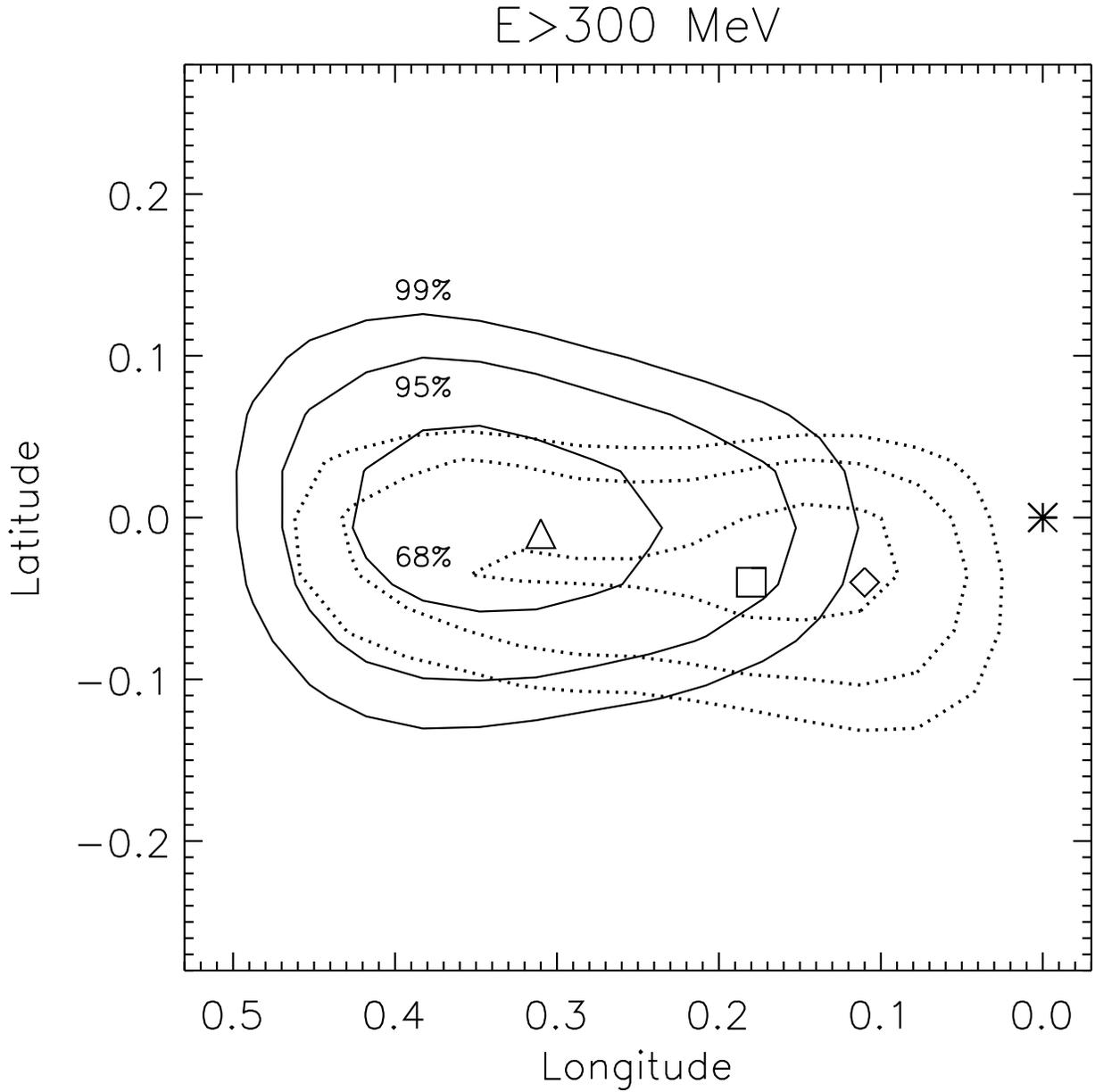}
\caption{The results of the source localization analysis for the energy band $>$~300~MeV. The
solid lines denote the confidence contours derived from a likelihood analysis of the data in
the $>$~300~MeV band, whereas the dotted lines are obtained by adding the \TS fine maps 
that result from likelihood analysis of the statistically independent data in the 300--1.000 MeV
and $>$ 1.000 MeV bands. The meaning of the
symbols is the same as in Fig.~\ref{f3}.}
\label{f5}
\end{figure}

The test statistic, TS, of two statistically independent data sets is additive. We can 
therefore co-add the \TS fine maps for the energy 
bands 300--1.000 MeV and $>$ 1.000 MeV to obtain a \TS fine map for the total energy range $>$~300~MeV.
The localization information thus derived should be compatible with the results of a
likelihood analysis of the data in the band $>$~300~MeV, which is systematically different for
two reasons.
First it lacks the information which photons have a high energy and thus a narrow PSF,
and therefore there is less information available. Second, the source spectrum may be different from
the simple power law that is assumed to construct a band-averaged PSF, and therefore the
PSF used in the likelihood analysis is the more likely to be inaccurate the wider
the energy band. In Fig.~\ref{f5} we compare the confidence contours derived from a likelihood 
analysis of the data in the $>$~300~MeV band with those obtained by adding the \TS fine maps
for the two narrower bands. There is a substantial overlap of the confidence regions, and 
the differences can be attributed to the displacement of the source localization at
energies $>$ 1.000 MeV relative to that at lower energies, an information that is preserved in the
co-added \TS fine map but not available to the likelihood analysis for the energy band $>$~300~MeV.
We do not note any systematic problem that might arise from a possibly inaccurate PSF in the
analysis of the wide energy band.
Again, the position of Sgr A$^\ast$ and the position of the TeV \gr source seem to be excluded,
and the GeV \gr source \gcs appears to be located a fraction of a degree away from the exact Galactic
Center.

\subsubsection{Consistency of localization in individual viewing periods}
The localization accuracy obtained from the analysis of data for individual viewing periods
will be worse than that derived using the combined data set, and the best-fit positions of \gcs
will be scattered over an area much larger than the confidence regions shown in Figs.~\ref{f3}--\ref{f5}.
We can test whether or not the observed scatter in the best-fit source position is compatible with
a common true position and the measured localization uncertainties. The confidence contours can usually
be accurately fitted with an ellipse that is specified by a semimajor axis $p$, semiminor axis $q$,
and position angle of the major axis $\phi$ \citep{th95}. The ellipse is centered on 
the centroid of the confidence location region, which thus serves as a position estimate. 
\citet{mat97} noted that for strong sources the dependence
of the logarithm of the likelihood of EGRET data upon the assumed source position closely 
follows a paraboloid, indicating a Gaussian error distribution and thus justifying the use of
the $\chi^2$-test. In contrast to \citet{th95} and \citet{mat97} we use a fit to the 68\% confidence 
contour. The displacement of the assumed (or measured) position from the true position is
described by the distance $r$ and the position angle $\theta$. Then the $\chi^2$-sum is defined by
\be
\chi^2 = \sum_{i=1}^N \,r_i^2\,\left[\left({{\cos(\phi_i-\theta_i)}\over {p_i^2}}\right)^2 +
\left({{\sin(\phi_i-\theta_i)}\over {q_i^2}}\right)^2\right]
\label{29}
\ee
A certain minimum in detection significance is necessary to provide a meaningful localization. Here
we only consider data from viewing periods for which \gcs was detected with at least $\sqrt{\TS}=3$.

\begin{deluxetable}{rrrrrr}
\tablecolumns{6}
\tablewidth{0pc} 
\tablecaption{Results of the localization analysis in individual viewing periods and the energy
band $>$~300~MeV. The position angle $\phi=0$ and $\phi=\pi/2$ correspond to the 
direction of the galactic-longitude axis and the negative galactic-latitude axis, respectively.
We only list viewing periods for which \gcs was detected with at least $\sqrt{\TS}=3$.
The best-fit positions are in degrees, the ellipse axes are in 
arcmin, and the major-axis position angle is in degrees.}
\tablehead{\colhead{ }&\multicolumn{2}{|c|}{position}&\multicolumn{3}{|c|}{68\% error ellipse}\\
\colhead{VP}&\colhead{$l$}&\colhead{$b$}&\colhead{$p$}&\colhead{$q$}
&\colhead{$\phi$}}
\startdata
5&0.05&0.21&22.2&7.9&93.0\\
16&0.39&-0.04&9.5&9.5&0.0\\
232&0.11&0.17&15.8&14.9&140.7\\
323&0.58&-0.04&15.9&15.9&0.0\\
324&0.48&-0.04&39.2&26.7&93.3\\
330&0.79&-0.38&26.6&21.1&120.9\\
332&0.27&-0.04&25.0&14.2&109.1\\
422&0.11&0.33&39.7&28.4&177.2\\
423&0.40&-0.33&47.8&23.0&13.1\\
429&0.81&-0.04&23.8&23.8&0.0\\
508&0.11&-0.04&22.4&22.4&0.0\\
625&0.31&-0.24&43.2&20.8&132.7\\
all&0.31&-0.01&7.3&3.5&176.3\\
\enddata
\label{t4}
\end{deluxetable}
Table~\ref{t4} lists the results of the location analysis in the energy range $>$~300~MeV
for the 12 viewing periods, for which 
the detection significance exceeded the required threshold. By minimizing the $\chi^2$-sum as given 
in Eq.~\ref{29} we obtain for that energy range
\be
\chi_{\rm min}^2=8.876\qquad N_{\rm d.o.f.}=10\qquad P_{\rm chance}=54.4\%
\label{30}
\ee
The coordinates, for which the $\chi^2$-sum is minimized, and their marginalized 1$\sigma$-errors are
\be
l=0.38^\circ\pm 0.09^\circ \qquad b=-0.007^\circ\pm 0.075^\circ
\label{31}
\ee
which is right inside the 68\% localization contour derived from teh combined data set for the same 
energy band (Fig.~\ref{f5}).
The scatter in the position estimates for the 12 viewing periods and the best-fit
position thus determined are therefore entirely consistent
with the statistical uncertainties. There is no evidence for an additional systematic error in the 
localization.

\section{Summary and discussion}

In this paper I have re-analyzed EGRET data of the observing periods 1--4 to derive the variability properties 
and localization of \gcsn. For that purpose I have introduced corrections in the way the diffuse foreground
model is used treated in the likelihood analysis, and I have also performed various consistency
checks using data for different energy bands and from different observing dates. Using only data for which I found no
evidence of systematic problems, e.g. excluding the wide energy band 100--10.000~MeV
and other low-energy bands, I obtained the following results:

\begin{itemize}

\item There is little evidence of variability. The chance probability
of drawing the measured distribution of photon fluxes given a constant flux is always higher than 
10\%, corresponding to a $\chi^2/$d.o.f. smaller than 1.35. The observed scatter in the measured 
flux values may correspond to variability at a level of 20--30\%, if it were real. Thus I
find an upper limit of $\lesssim 30\%$ to the relative amplitude of stochastic
variability in the lightcurve above 300 MeV of \gcsn.
This conclusion would be corroborated by the data in the energy range 100--300 MeV, if I
neglected viewing period 429, for which one notes an apparently intense flare with a
soft spectrum. I find some indications of systematic problems in the data of VP429 for the 100--300 MeV
band, but no substantial indication that the observed high flux may be an artifact.

\item The data indicate that \gcs is displaced from the exact Galactic Center towards positive Galactic longitudes,
in qualitative agreement with the findings of \citet{hd02}. This displacement is consistently found in 
maximum likelihood analyses of all energy bands above 300~MeV and also using data of individual viewing periods
and the energy range 300--10.000~MeV, in contrast to the contradictory results of \citet{hmh98}. 
The position of Sgr A$^\ast$, the center of Sgr A East, the pulsar J1747-2958,
and the position of the TeV \gr source seem to be incompatible at the $> 95\%$ level with the
results of the localization analysis of \gcsn. The Galactic Center arc 
or any other source located in the
Galactic plane at longitudes $0.1^\circ \lesssim l\lesssim 0.4^\circ$ are possible counterparts to the GeV \gr source.
On account of the similarity of the spectra of \gcs and that of the diffuse Galactic \gr emission
it is possible, though very unlikely, that a compact, dense complex of interstellar gas, that is unaccounted 
for in the standard foreground model, is responsible for \gcsn. The total mass of the gas cloud
would have to be $M_{\rm gas} \simeq 5\cdot 10^7\,M_\odot$ if located at the Galactic Center, 
and the cloud should have an apparent extent of $\theta_{\rm gas} \lesssim 0.3^\circ$, lest it doesn't 
appear point-like at high \gr energies. Such a massive
gas complex should have been seen in the atomic fine-structure lines \citep{oiha01,rod04} or in CO lines 
\citep{kim02,mar04}. A somewhat smaller mass of unaccounted gas is required to shift the apparent position
of a true point source, but even at the 20\% level, i.e. $M_{\rm gas} \simeq 10^7\,M_\odot$, the radio and infrared
data do not indicate the existence of a previously unknown gas of the appropriate localization and compactness.

\end{itemize}

\section{Acknowledgements}
Support by NASA under award No. NAG5-13559 is gratefully acknowledged. The author wishes to thank John Mattox 
for valuable advise on some intricacies of the LIKE software.


\begin{thebibliography}{}

\bibitem[Aharonian et al.(2004)]{aha04} Aharonian, F., et al., 2004, \aap, 425, L13

\bibitem[Atoyan \& Dermer(2004)]{ad04} Atoyan A., and Dermer C.D., 2004, \apj, 617, L123

\bibitem[Berezinsky et al.(1992)]{bere92} Berezinsky V.S., Gurevich A.V., Zybin K.P., 1992,
Phys. Lett., B294, 221

\bibitem[Bergstr\"om et al.(1998)]{ber98} Bergstr\"om L., Ullio P., Buckley J.H., 1998, APh, 9, 137

\bibitem[Crocker et al.(2004)]{cro04} Crocker R.M., Fattuzzo M., Jokipii R., et al., \apj, submitted
(astro-ph/0408183)

\bibitem[Ellis et al.(2002)]{ell02} Ellis J., Olive K.A., Santoso Y., Spanos V.C., 2002, Phys.Let. B, 565, 176

\bibitem[Esposito et al.(1999)]{esp99} Esposito J.A., et al., 1999, \apjs, 123, 203

\bibitem[Fattuzzo \& Melia(2003)]{fm03} Fattuzzo M., Melia F., 2003, \apj, 596, 1035

\bibitem[Gnedin \& Primack(2004)]{gp03} Gnedin O.Y., and Primack J.R., 2003, \prl, 93, 061302 

\bibitem[Hartman et al.(1999)]{hart99} Hartman R.C., et al., 1999, \apjs, 123, 79

\bibitem[Hooper \& Dingus(2002)]{hd02} Hooper D. and Dingus B., 2002, preprint (astro-ph/0212509)

\bibitem[Hunter et al.(1997)]{hunter97} Hunter S.D., et al., 1997, \apj, 481, 205

\bibitem[Kime et al.(2002)]{kim02} Kim S., Martin C., Stark A., Lane A., 2002, \apj, 580, 896

\bibitem[Kosack et al.(2004)]{kos04} Kosack K., Badran H.M., Bond H.I., et al., 2004, \apj, 608, L97

\bibitem[Liu et al.(2003)]{lpm04} Liu S., Petrosian V., Melia F., 2004, \apj, 611, L101

\bibitem[Mahadevan et al.(1997)]{maha97} Mahadevan R., Narayan R., Krolik J., 1997, \apj, 486, 268

\bibitem[Markoff, Melia and Sarcevic(1997)]{mms97} Markoff S., Melia F., Sarcevic I., 1997, \apj, 489, L47

\bibitem[Martin et al.(2004)]{mar04} Martin C., Walsh W., Xiao K. et al., 2004, \apjs, 150, 239

\bibitem[Mattox et al.(1997)]{mat97} Mattox J.R., et al., 1997, \apj, 481, 95

\bibitem[Mattox et al.(1996)]{mattox96} Mattox J.R., et al., 1996, \apj, 461, 396

\bibitem[Mayer-Hasselwander et al.(1998)]{hmh98} Mayer-Hasselwander, H.A., et al., 
1998, \aap, 335, 161

\bibitem[McLaughlin \& Cordes(2003)]{mc03} McLaughlin M.A., Cordes J.M., 2004, 
Proceedings of the 4th AGILE Science Workshop on 
{\it X-ray and Gamma-ray Astrophysics of Galactic Sources}, in press, astro-ph/0310748

\bibitem[Melia \& Falcke(2001)]{mf01} Melia F., Falcke H., 2001, \araa, 39, 309

\bibitem[Merck et al.(1996)]{mer96} Merck M., et al., 1996, \aaps, 120, C465

\bibitem[Nolan et al.(2003)]{nol03} Nolan P.L., et al., 2003, \apj, 597, 615

\bibitem[Ojha et al.(2001)]{oiha01} Ojha R., Stark A., Hsieh H., et al., 2001, \apj, 548, 253

\bibitem[Pohl(1997)]{po97} Pohl M., 1997, \aap, 317, 441

\bibitem[Rodr\'iguez-Fern\'andez et al.(2004)]{rod04} Rodr\'iguez-Fern\'andez N.J., Mart\'in-Pintado J., Fuente A., 
Wilson T.L., 2004, \aap, 427, 217

\bibitem[Sch\"odel et al.(2002)]{sch02} Sch\"odel R.R., Ott T., Genzel R., et al., Nature, 419, 694

\bibitem[Thompson et al.(1995)]{th95} Thompson, D.L., et al., 1995, \apjs, 101, 259

\bibitem[Tsuchiya et al.(2004)]{tsu04} Tsuchiya K., Enomoto R., Ksenofontov L.T., et al., 2004, \apj,
606, L115

\bibitem[Wilks(1938)]{wilks} Wilks, S.S., 1938, Ann. Math. Stat., 9, 60

\end{thebibliography}
\end{document}